\documentclass[]{aa}

\usepackage{graphicx}
\usepackage{natbib}   
\usepackage{scalerel}
\usepackage{longtable}
\usepackage{bm}
\usepackage{soul}
\usepackage{amsmath}
\usepackage[normalem]{ulem}
\usepackage[dvipsnames]{xcolor}
\usepackage{pifont}

\newcommand{\nside}{\ensuremath{N_{\textrm{side}}}}

\usepackage[pdfencoding=auto,psdextra]{hyperref}
\hypersetup{
    colorlinks=true,
    linkcolor=blue,
    filecolor=magenta,
    urlcolor=blue,
    citecolor=blue
}
\usepackage[nameinlink]{cleveref}

\usepackage{eso-pic}

\AddToShipoutPictureBG*{
  \AtPageUpperLeft{
    \hspace{\paperwidth}
    \raisebox{-2.8cm}{
      \makebox[0pt][r]{IFT-UAM/CSIC-26-036 \hspace{1.5cm}}
    }
  }
}

\newcommand\blfootnote[1]{%
\begingroup
\renewcommand\thefootnote{}\footnote{$^\dagger$ #1}%
\addtocounter{footnote}{-1}%
\endgroup
}

\begin{document}
\title{More Than Power:\\Revisiting the CMB Hemispherical Power Asymmetry\\with Morphological Descriptors}
\author{Javier Carrón Duque\inst{1}\,\thanks{Email: \href{mailto:javier.carron@csic.es}{javier.carron@csic.es}} \and Mikel Martin Barandiaran\inst{1,2}\,\thanks{Email: \href{mailto:mikel.martin@uam.es}{mikel.martin@uam.es}} \and Joseba Martínez-Arrizabalaga\inst{3,2}\,$^\dagger$ }

\institute{$^{1}$Instituto de Física Teórica UAM-CSIC, Campus de Cantoblanco, 28049 Madrid, Spain\\
$^{2}$Departamento de Física Teórica, Universidad Autónoma de Madrid, 28049 Madrid, Spain \\ $^{3}$Departamento de Electrónica y Electromagnetismo, Facultad de Física, Universidad de Sevilla, 41012 Sevilla, Spain}

\abstract
   {The Cosmological Principle assumes a statistically isotropic Universe, but the Cosmic Microwave Background (CMB) exhibits some anomalous statistical features, such as the hemispherical power asymmetry, that challenge this core assumption.}
   {We aim to expand the characterization of this anomaly by investigating the uniformity of the CMB's morphological and topological properties, testing whether the asymmetry extends beyond the variance of the temperature field.} 
   {We evaluate the three Minkowski Functionals (MFs) on local patches of the \textit{Planck} SMICA temperature map and compare them against 999 FFP10 realistic simulations including noise. By fitting the local MFs to the analytical expectations for Gaussian isotropic fields, we extract local estimates of the temperature variance ($\sigma^2$), the variance of the field gradients ($\tau^2$), and the goodness-of-fit ($\chi^2$). We then evaluate the presence, amplitude, and alignment of dipoles in these locally varying quantities.}
   {We confirm the highly significant dipole in $\sigma^2$ (p-value $\sim0.3\%-1.0\%$). Furthermore, we report a moderately significant dipole in the gradient variance $\tau^2$ (p-value $\sim2.6\%-3.7\%$) that is statistically independent of the $\sigma^2$ dipole under the $\Lambda$CDM model. We also find a mild spatial variation in the goodness-of-fit to the Gaussian isotropic predictions (p-value $\sim2.1\%-5.6\%$). Remarkably, the \textit{Planck} dipoles for all three quantities are clustered and point toward the same region of the sky. We find that the asymmetries in all quantities are well described by dipoles.}
   {The Hemispherical Asymmetry in the CMB extends beyond its local variance, as we found it to be also present in its local morphology. The MFs of the CMB present a dipolar pattern that can be characterized by independent dipoles in the variance and gradient variance, plus a hint of a non-Gaussian nature of this asymmetry. These results provide a more complete characterization of the Hemispherical Asymmetry, and will therefore contribute to better determine the nature of the physical effect behind it, whether cosmological, residual foregrounds, or unknown systematics.
}
   
\keywords{Cosmology: theory -- (Cosmology:) cosmic background radiation -- Surveys -- Methods: statistical -- Methods: data analysis}

\maketitle

\section{Introduction}
The Cosmological Principle is one of the core assumptions of modern cosmology, being a cornerstone of the $\Lambda$CDM model and many of its extensions. It states that the universe is homogeneous and isotropic on sufficiently large scales. This assumption, initially established from a philosophical point of view, became empirically testable at the end of the twentieth century thanks to the discovery of the Cosmic Microwave Background (CMB) perturbations. Until the present day, three full-sky microwave missions have been launched, aiming to measure the CMB and its anisotropies with increasing precision: COBE \citep{Smoot:1991pb}, WMAP \citep{WMAP:2003elm}, and \textit{Planck} \citep{Planck:2013oqw}.
\blfootnote{All authors contributed equally to this work. The ordering of the authors is merely alphabetical.}

Although the CMB data is roughly compatible with the Cosmological Principle, many researchers have claimed that the CMB reveals a subtle but statistically significant violation of this principle, as several different statistics are found to be in mild tension with the assumption of statistical isotropy of the CMB; these are called the \emph{CMB anomalies} \citep[see, e.g, ][]{Tegmark:2003ve, deOliveira-Costa:2003utu, Vielva:2003et, Larson:2004vm, Land:2004bs, Land:2005ad, Land:2006bn, McEwen:2004sv, McEwen:2006yc, Jaffe:2006sq, WMAP:2006jqi, WMAP:2006bqn, Cruz:2006fy, Bridges:2006mt, Copi:2006tu, Bernui:2005pz, Bernui:2008ve, Pietrobon:2008rf,%
Axelsson:2013mva,Mukherjee:2015wra,Billi:2019vvg,Billi:2023liq,Natale:2019dqm,Galloni:2022rgg,Yeung:2022smn,Aluri:2022hzs,Shi:2022hxc,Galloni:2023pie, Sanyal:2024iyv,Gimeno-Amo:2025icf}. Four representative statistics of the CMB anomalies have recently been shown to be independent, even in the tails, under the (isotropic) $\Lambda$CDM model, implying that the violation of the Cosmological Principle may be much more significant than any of these anomalies independently suggest \citep{Jones:2023ncn}.

Among the CMB anomalies, one of the most significant is the so-called \emph{hemispherical power asymmetry}: a directional dependence of the variance of the CMB, \textit{i.e.}, its power \citep{Eriksen:2003db}. This directional dependence can be characterised as a dipole pointing at $(\ell,\,b) \approx (225^\circ, \, -27^\circ)$ in the variance of the map \citep{Eriksen:2007pc}. Thereafter, multiple works have reported the presence of this asymmetry in both WMAP and \textit{Planck} data with different methodologies, reporting significances around $3\sigma$ \citep{Hansen:2004vq, Hajian:2004zn, Paci:2010wp, Rath:2013yra, Akrami:2014eta, Aiola:2015rqa, Ghosh:2015qta, Shaikh:2019dvb, Gimeno-Amo:2023jgv, Kester:2023qmm, Sanyal:2024iyv, Sanyal:2026cvk}. Given that this anomaly is explicitly connected to a violation of statistical isotropy, a deeper characterization and understanding of this anomaly could open the door to a more complete understanding of the validity of the Cosmological Principle. 

The mechanism responsible for the power asymmetry is not yet understood, but several possibilities have been proposed in the literature. One possibility is that it is caused by some residual effect of galactic foregrounds or unknown systematics \citep{Rassat:2014yna}. Another possibility is that it is a secondary CMB effect caused by the nearby Universe, possibly through the Integrated Sachs-Wolfe effect of large structures \citep{Inoue:2006rd, Bernui:2009pv, Naidoo:2017woy} or some possibly unknown physical effect related to nearby galaxies \citep{Hansen:2023gra, Hansen:2024vgs}. A third possibility is that the early Universe really is anisotropic, with different models being proposed in the literature, including causal horizons \citep{Fosalba:2020gls} or modified power spectra, possibly as a result of a new primordial field acting as a modulating field \citep{Watanabe:2010bu,Soda:2012zm,McDonald:2014lea,Gaztanaga:2024vtr, Gandhi:2025wxb}. A particularly interesting option is that the Universe could have a non-trivial spatial topology \citep[][and references therein]{COMPACT:2025adc,Samandar:2025kuf,COMPACT:2024qni}: this would introduce a violation of statistical isotropy in the Universe without the need to extend neither General Relativity nor the Standard Model of Particle Physics, but it has not been shown yet that the anisotropies induced are the kind needed to explain the CMB anomalies \citep{COMPACT:2026ano}. Although this list is far from exhaustive, it is clear that we need more insights on the hemispherical power asymmetry in order to potentially distinguish between different proposals.

In this work, we explore a novel aspect on the hemispherical asymmetry by analysing the uniformity of morphological descriptors of the CMB. In particular, we study the three Minkowski Functionals (MFs), a popular set of statistics that characterize the morphology and topology of a field with a rich body of research in the Mathematical literature\footnote{In the Mathematical literature they are sometimes referred to as Lipschitz-Killing curvatures. They differ from MFs in the ordering of the labels and constant multiplicative factors.} \citep{adler2007,kratz2018central,vidotto2022random,cammarota2023lipschitz,CarronDuque:2024qxl,Lerario:2025aol}. MFs are sensitive to both the variance of the map, $\sigma^2$, and the variance of the gradient of the map, $\tau^2$. In fact, for zero-mean Gaussian isotropic fields, these two quantities are the only ones that come into play in the theoretical expected value of the MFs. The fact that this expected value is known with low variance, together with the fact that the CMB is at least nearly Gaussian and isotropic, make MFs a powerful tool to study the hemispherical asymmetry. In this work, we study whether the MFs of the CMB are uniform over the sky and, if not, how this relates to a dipole in $\sigma^2$, as well as assessing whether there are additional dipoles in $\tau^2$ and in the goodness-of-fit to the Gaussian prediction. By exploring these complementary statistics (i.e., by looking elsewhere), we reduce the reliance on the metrics where the anomaly was initially found. This broader perspective helps us understand the asymmetry better, with the goal of distinguishing between its proposed origins.

This paper is structured as follows. In \Cref{sec:MFs} we revise the mathematical foundations behind MFs, paying special attention to their predictions for isotropic Gaussian random fields on the sphere. In \Cref{sec:data} we detail the data and simulations that we studied, together with the methodology that we implemented. In \Cref{sec:results} we present our results characterizing the hemispherical asymmetry. Finally, in \Cref{sec:conclusions} we summarize our conclusions and broader implications.

\section{Minkowski Functionals}\label{sec:MFs}

MFs describe the geometry and morphology of a field. They were introduced in the context of stochastic geometry and have been applied to many fields, such as medical imaging \citep{Nagarajan:2014amf,bierme2019lipschitz}, material science \citep{Scholz_2012}, or neural networks \citep{di2025fractal}. Cosmology has also adopted these functionals to study the properties of cosmological fields, such as CMB observables \citep{Winitzki:1997jj,Schmalzing:1997uc,Planck:2015igc,Planck:2019evm,martire2023morphological,Chingangbam:2023moo,Ranucci:2025lao,vacher2025bad,campeti2025few,puglisi2026extending}, large-scale structure of the Universe \citep{Hikage:2002ki,Wiegand:2016ezl,Sullivan:2017mhr,Appleby:2021lfq,Appleby:2022itn,Liu:2025haj}, weak lensing \citep{Parroni:2019snd,Euclid:2023uha,Hamann:2023tdu,armijo2025cosmological,hoyland2025fast}, or 21cm intensity maps \citep{Kapahtia:2021eok,Spina:2021drb}.

MFs are defined as follows. First, we take a field $f$ defined on some $d$-dimensional space $\mathcal{M}$ (e.g., CMB temperature on the sphere, galaxy overdensity on a survey volume,...) and we set a threshold $u\in\mathbb{R}$ (e.g., zero, two times the standard deviation of the field,...). We define the excursion set of $f$ at $u$, called $A_f(u)$, as the regions of the space where the field takes a value of $u$ or higher:
\begin{equation}
    A_{f}(u) = \{ x \in\mathcal{M} : f(x)\geq u  \}.
\end{equation}
If the space $\mathcal{M}$ is $d$-dimensional, we can define $d+1$ MFs as in \cite{Schmalzing:1997uc}:
\begin{equation}
    V_0(u) = \int_{A_f(u)} {\rm d}V\,,
    \label{eq:MF_integral1}
\end{equation}
\begin{equation}
    V_j(u) = \frac{1}{s_{j-1}\binom{d}{j}}\int_{\partial A_{f}(u)} \mathcal{\vartheta}_j(\kappa_1, \dots ,\kappa_{d-1})\, {\rm d}S \,
\end{equation}
where ${\rm d}V$ and ${\rm d}S$ are the volume and surface elements of the excursion set and its boundary, $s_{j-1}$ is the (hyper-)area of the $j$-sphere, $\mathcal{\vartheta}$ is the symmetric polynomial of order $j$, and $\kappa_1, \dots ,\kappa_{d-1}$ are the principal curvatures of the surface boundary of the excursion set, see \cite{Schmalzing:1997uc,adler2007} for more details.

In this work, we focus on the CMB temperature, so we can particularize the formalism to a two-dimensional field defined on the sphere:
\begin{align}
    V_0(u) &= \int_{\mathbb S^2} \Theta(f(\hat{n}) - u)\,  {\rm d} \hat{n}\,,  \\
    V_1(u) &= \frac{1}{4}\int_{\partial A_{f} (u)}{\rm d} \ell\,, \\
    V_2(u) &= \frac{1}{2\pi}\int_{\partial A_{f} (u)} \kappa \, \mathrm{d}\ell\,,
\end{align}
where $\Theta$ is the Heaviside step function and the integrals are done with respect of the area element on the sphere $\mathrm{d} \hat{n}$, or the length element of the boundary length $\mathrm{d}\ell$.

The first MF, $V_0$, is straight-forwardly interpreted as the total volume of the excursion set. The second MF, $V_1$, corresponds to the so-called \emph{boundary length}, i.e., the total perimeter of the excursion set. Lastly, the third MF, $V_2$, can be interpreted as the Euler-Poincaré characteristic of the excursion set, sometimes denoted $\chi(A_{f}(u))$. This last interpretation is due to the Gauss-Bonnet theorem, which relates the curvature on the boundary of a set (a geometric quantity) with the Euler-Poincarè characteristic of said set (a topological quantity). MFs can therefore describe both geometric and topological characteristics of a given field as a function of the threshold $u$. Additionally, the Hadwiger theorem \citep{hadwiger1975,chen2004} states that any other descriptor of the excursion set that satisfies some desirable conditions (more technically, being a valuation) can be written as a linear combination of the MFs. This makes MFs a very general tool to study the morphology of fields.

When applied to a random field, the MFs become powerful statistics as a function of the threshold $u$. There is a rich body of literature studying the statistical properties of MFs when applied to different types of random fields, the most important to us being Gaussian isotropic fields \citep{tomita1986,Schmalzing:1997uc}. Additionally, there is a significant amount of literature on other fields, including weakly non-Gaussian fields \citep{Hikage:2006fe,Codis:2013exa,Matsubara:2020knr}, some types of anisotropic Gaussian fields \citep{bierme2025anisotropy}, quadratic and cubic fields \citep{Fantaye:2014vua}, spin fields \citep{stecconi2021,Lerario:2025aol,pistolato2025,Pistolato:2024sdn}, time-depending fields \citep{leonenko2025high,ruiz2025testing}, or fractals \citep{zaehle2010,rataj2022}.

In this work, we focus on the prediction for Gaussian isotropic fields defined on the sphere, as the CMB is known to be very close to such a field. In this case, the expectation value of the MFs can be computed analytically as a function of only two parameters: the variance of the field $\sigma^2$, and the variance of the modulus of its gradient $\tau^2$:

\begin{equation}
    \sigma^2 = \langle f(x)^2 \rangle = \sum_{\ell=2}^{\infty} \frac{2\ell +1}{4\pi} C_\ell\,,
    \label{eq:sigma2}
\end{equation}
\begin{equation}
    \tau^2 = \langle |\nabla f(x)|^2 \rangle = \sum_{\ell=2}^{\infty}  \frac{\ell(\ell+1)}{2} \frac{2\ell +1}{4\pi} C_\ell\,,
    \label{eq:tau2}
\end{equation}
where $C_\ell$ is the angular power spectrum of the field $f$. These two quantities fully characterise the underlying morphology of a Gaussian isotropic field. 
The expected value of the MFs for a zero-mean Gaussian isotropic field can then be written as:
\begin{equation}
    \frac{\mathbb{E}\left[V_{0}(u)\right]}{4\pi }=1-\Phi \left(\frac{u}{\sigma}\right)\,,
    \label{eq:MF0_gaussian}
\end{equation}
\begin{equation}
    \frac{\mathbb{E}\left[V_{1}(u)\right]}{4\pi }=\frac{1}{8}\exp \left(-\frac{u^{2}}{2\sigma^2}\right) \frac{\tau}{\sigma} \,,
    \label{eq:MF1_gaussian}
\end{equation}
\begin{equation}
    \frac{\mathbb{E}\left[V_{2}(u)\right]}{4\pi }=
\frac{1}{\sqrt{(2\pi )^{3}}}\, \frac{\tau^2}{\sigma^2} \exp \left(-\frac{u^{2}}{2\sigma^2}\right)\frac{u}{\sigma}\,,
    \label{eq:MF2_gaussian}    
\end{equation}
where $\Phi$ is the cumulative distribution function of the standard Gaussian distribution. Notice that $\mathbb{E}(V_i)\propto(\tau/\sigma)^{i}$; this ratio is sometimes called $\mu$, although we will not use that notation in this work. From these predictions, we can see that $\sigma$ acts as a scaling factor in the threshold $u$ (and therefore a horizontal stretching of the MF curves), while $\tau / \sigma$ acts as a scaling factor in the amplitude of the MFs (a vertical stretching of the MF curves). One can invert these relations to obtain the best value of $\sigma$ and $\tau$ that describe the morphology of a given (Gaussian isotropic) field by fitting the measured MFs to these expressions. We apply these formulae to local patches rather than the full sky. Given that all hints of violation of statistical isotropy seem to be confined to the large scales, we expect these formulae to hold on the aforementioned patches. We shall verify that our results are consistent with this expectation.

\section{Data and Methodology}\label{sec:data}
\begin{figure}
\includegraphics[width=\linewidth]{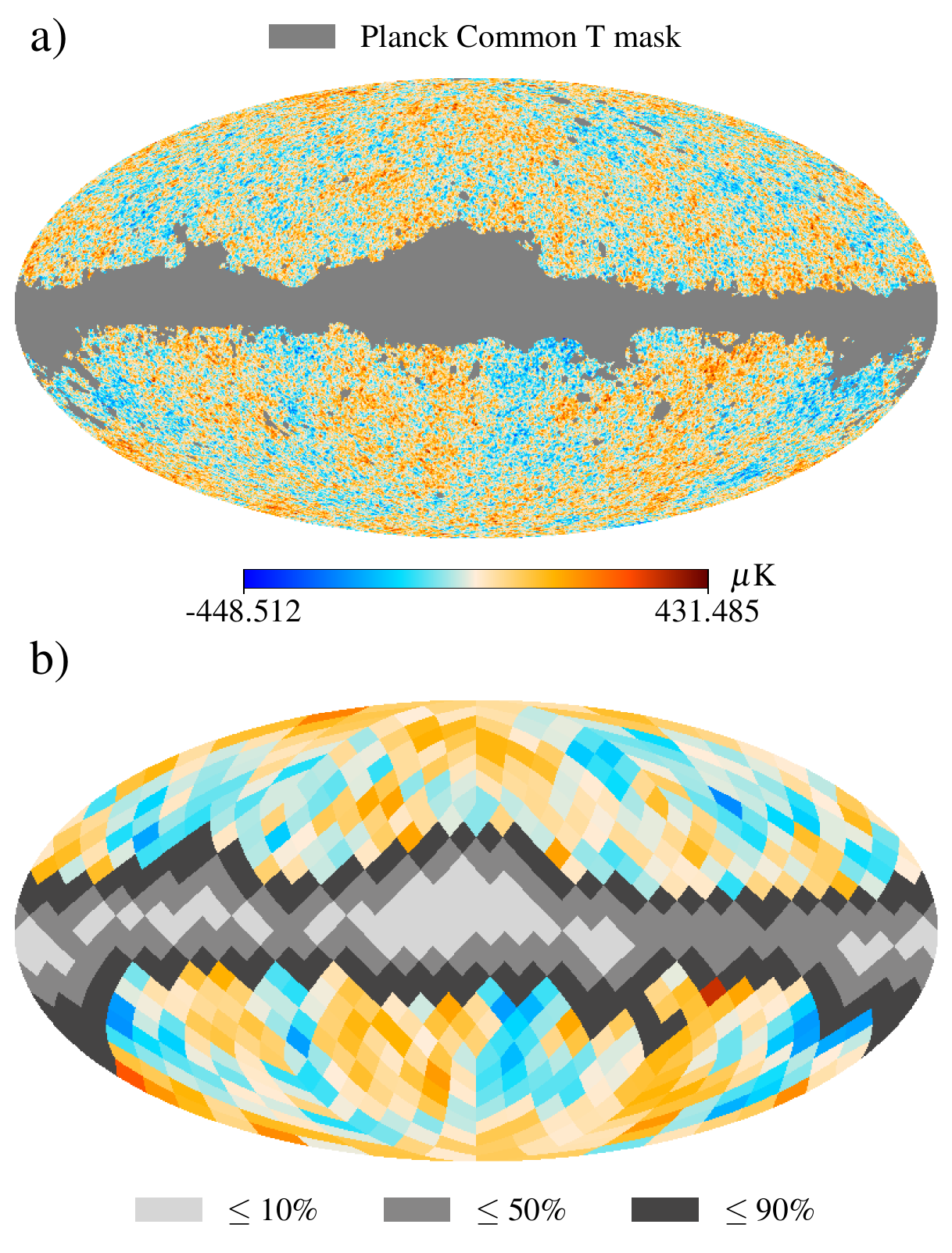}
\caption{\textit{Top panel (a):} The \textit{Planck} SMICA temperature map used as the baseline for our analysis. The gray region illustrates the \textit{Planck} common temperature mask, which excludes diffuse foreground emission and point sources. \textit{Bottom panel (b):} The distribution of the 768 local analysis patches, centered on an $N_{\rm side} = 8$ grid. Colored pixels represent valid disc locations. Grayscale pixels indicate disc centers that are excluded based on the fraction of unmasked pixels they contain. The different shades correspond to varying stringency thresholds ($\leq10\%$, $\leq50\%$, and $\leq90\%$ valid pixels). Our baseline analysis excludes regions falling under the 50$\%$ criterion (light and medium gray).}
\label{fig:masks}
\end{figure}

In this work we analyse the uniformity of the morphology of the CMB. We use the last official data release by the \textit{Planck} Collaboration, DR3 \citep{Planck:2018nkj}. Among the four different reconstruction methods, we have chosen to analyse the maps produced by the SMICA pipeline \citep{Cardoso:2008qt, Planck:2018yye}. To evaluate the significance of our results, we compare them with the Full Focal Plane 10 (FFP10) simulations of CMB and noise \citep{Planck:2019evm}, consistently propagated through the SMICA pipeline. Due to the computational difficulty of such simulations, there are only 300 noise realisations available, while there are 999 processed CMB realizations. Given that noise is sub-dominant, we decide to use the full set of CMB realisations, pairing each noise simulation to several CMB simulations as needed. In order to exclude pixels containing diffuse foreground emission and point sources, as well as to provide a proper comparison with Planck data, we have used the \textit{Planck} common temperature mask in all analyses. The resulting maps have 77.9\% sky coverage, as shown in the top panel of \Cref{fig:masks}. Ideally, we would have liked to repeat the analysis for all four reconstruction methods, as well as newer maps produced by reprocessing Planck data \citep{Planck:2020olo,Tristram:2023haj,Nofi:2025qkb}, but we had to limit the analysis to a single set of data and simulations due to computational constraints.

The processing of each map, which we describe in the following, is similar to \citet{Akrami:2014eta}, although with some differences. All data and simulations are provided as HEALPix\footnote{\url{https://healpix.sourceforge.io/}} maps \citep{2005ApJ...622..759G,Zonca2019}, with a native size of $\nside=2048$. In order to reduce the computational time to compute the MFs, and given that the hemispherical anomaly is a low-$\ell$ effect \citep{Akrami:2014eta,Gimeno-Amo:2023jgv,Sanyal:2026cvk},  we convolve all maps with a Gaussian kernel of $1^\circ$ and downsize them to $\nside=512$. We have verified that this step has no effect on the significance of the hemispherical asymmetry, while providing a significant reduction in the computational resources needed.

For each map, we consider 768 discs centred in the middle of the pixels of a HEALPix map of $N_{\rm side} = 8$, with a radius of 10$^\circ$. This choice ensures that every point on the map is covered by at least one disc, as the distance between disc centres is approximately $\sqrt{4\pi /N_{\rm pix}} \approx7.33^\circ$; we note that the results in each patch are not fully independent from each other, as there is some overlap between discs. As previously mentioned, we have considered the \textit{Planck} common temperature mask, shown in \Cref{fig:masks} (top panel). Additionally, we only use the discs that contain at least $50\%$ of valid pixels; we note that this choice is more conservative than the typical choice in the literature of using discs that contain at least $10\%$ of valid pixels. The discarded pixels in the $N_{\rm side} = 8$ maps according to different fractional criteria can be seen in the bottom panel of \Cref{fig:masks}. Our conservative choice rejects 162 out of 768 pixels, which amounts to $21\%$ of the map. 

We have verified that our results are robust to changes in the disc size between $8^\circ$ and $12^\circ$, as well as to changes in the minimum valid pixel fraction between $10\%$ and $90\%$. This includes changes in both choices simultaneously. Therefore, in the following we restrict our discussion to the baseline choices of $10^\circ$ disc radius and $50\%$ minimum valid pixel fraction.

\begin{figure}
\includegraphics[width=\linewidth]{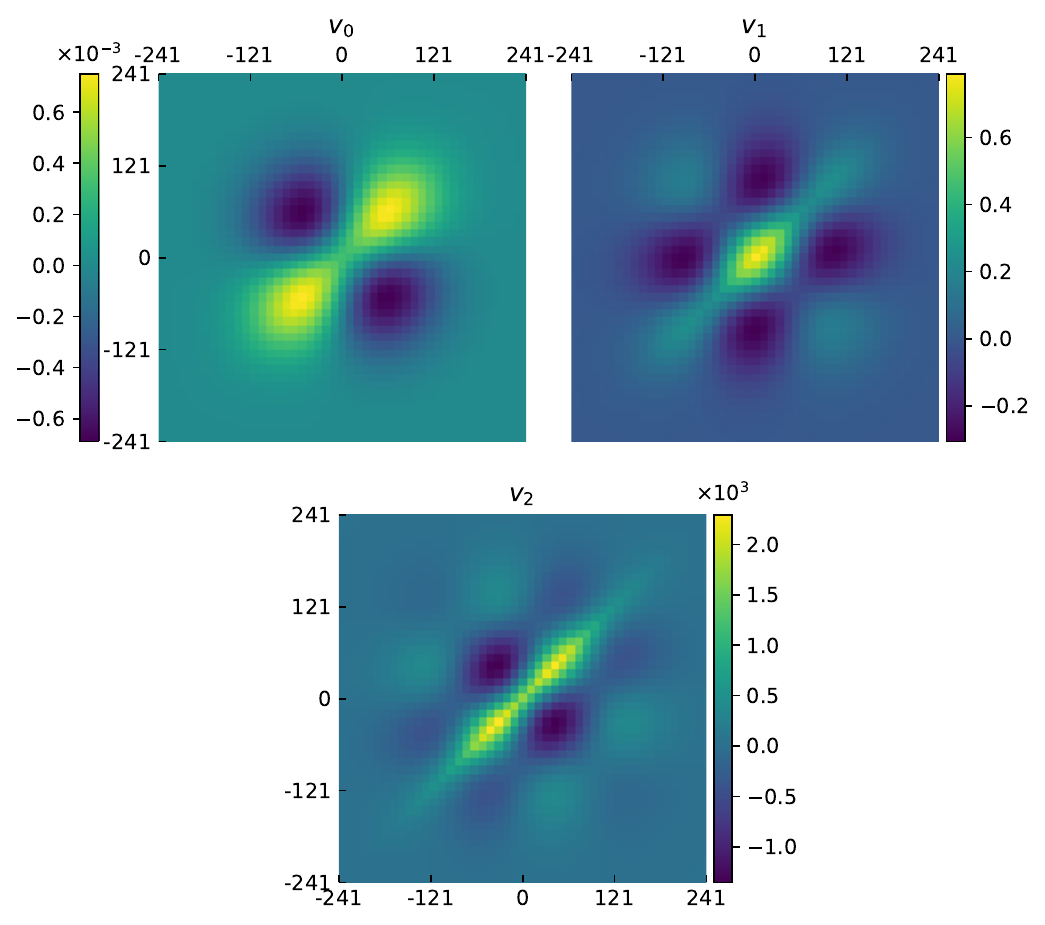}
\caption{Covariance matrices of the MFs for pairs of temperature thresholds between -241$\mu$K and $241\mu$K (approximately $-3.5\sigma$ to $3.5\sigma$).}
\label{fig:thresholdcov}
\end{figure}

Within each disc of each map, we calculate the three MFs using Pynkowski\footnote{\url{https://github.com/javicarron/pynkowski}} \citep{Carones:2022rbv,CarronDuque:2023bph}. 
We define a fixed range of thresholds, corresponding approximately to a range from $-3.5\sigma$ to $3.5\sigma$. Then, for each patch of each map, we offset these thresholds so that they are centred at the mean value of the patch, and evaluate the MFs at these values. We obtain three curves per disc and map, summing up to a total of $606000$ curves for each of the three MFs. We then compute the covariance between different thresholds for each MF independently, as can be seen in \Cref{fig:thresholdcov}; this covariance is computed using randomly selected non-overlapping discs. We fit each curve to the theoretical expectations in \Cref{eq:MF0_gaussian,eq:MF1_gaussian,eq:MF2_gaussian} using the \texttt{curve\_fit} function in scipy \citep{2020SciPy-NMeth}, estimating the best fit for $\sigma$ and $\tau$. Additionally, we compute the commonly-used \emph{patch variance} directly on each patch ($\sigma^2$), as well as the variance of the derivatives on the patch ($\tau^2$). Therefore, for a given patch, we obtain four estimates of $\sigma^2$ (one from each MF and one from the patch variance) and three estimates of $\tau^2$ (one from each MF except $V_0$, and one from the patch variance); these estimates should be consistent with each other if the field is Gaussian and isotropic, while the values for $\sigma^2$ and $\tau^2$ are uncorrelated, see \Cref{ap:corrs} for more details. Although in light of \Cref{eq:sigma2,eq:tau2} both $\sigma^2$ and $\tau^2$ are related to the angular power spectrum $C_{\ell}$, the different weightings in the sums imply that each quantity is sensitive to different scales: the former is affected mostly by larger scales $(\propto\ell\, C_\ell)$ whereas the latter is more sensitive to smaller scales $(\propto\ell^3\, C_\ell)$. Since power at small and large scales is assumed to be independent due to statistical isotropy, $\sigma^2$ and $\tau^2$ need not be correlated. We note that unbiased estimators for the location and scale of Gaussian fields from the MFs have been studied in the mathematical literature for fields in $\mathbb{R}^2$ \citep{di2022statistics}; to the best of our knowledge, the estimators for $\tau^2$ and for fields in the sphere have not been fully developed yet. 

As explained before, the values of the MFs for a Gaussian isotropic field are very well predicted. Therefore, we can compute how well this assumption holds for each of the computed MF on each patch. We do this through the goodness of fit for the best-fit parameters, which we approximate at first order as a $\chi^2$:
\begin{equation}
    \chi^2 = z^T\, C^{-1}\, z\,,
\label{eq:goodnessoffit}
\end{equation}
where $z = v_i(u) - \mathbb E \left[v_i(u)\right]$ is the residual between the measured and the expected value for the best-fit parameters, and $C^{-1}$ is the inverse covariance matrix, which is estimated as explained before, and corrected with the Hartlap factor \citep{Hartlap:2006kj}. In addition to the $\sigma^2$ and $\tau^2$ values, we therefore also obtain a value of $\chi^2$ for each MF in each patch and map (in this case there is no equivalent directly from the patch), which quantifies how well the morphology of the field in that patch is described by a Gaussian isotropic field with the best-fit parameters. We note that, unlike the other parameters, these $\chi^2$ values are not necessarily correlated between different MFs, as they quantify different aspects of the morphology of the field (i.e., a slightly non-Gaussian field may have a good fit in one MF but a bad fit in another).

After obtaining these values for each patch and map, we obtain $10\times1000$ maps: $3$ for $\sigma^2$, $4$ for $\tau^2$, and $3$ for $\chi^2$, times $999$ simulations and one Planck map. We then study their uniformity over the sky, and the presence of a possible dipole as in \citet{Akrami:2014eta,Adhikari:2014mua,Panda:2022tmt,Khan:2022lpx,Sanyal:2024iyv}. In order to this, we compute the mean field of each quantity as the average over all simulations. Then, for each map, including the Planck ones, we subtract this mean field and divide by it in order to remove possible systematic effects induced by the foregrounds or masks. Finally, we compute the dipole (and higher multipoles) of this ``contrast'' maps. 

In the next section we will present the results of this analysis, comparing the amplitude of the dipole in Planck data to the one obtained in simulations in order to assess its significance.

\section{Results}\label{sec:results}
We apply the methodology described in the previous section to the FFP10 simulations, as well as the Planck CMB temperature map, both consistently processed with the SMICA map-making algorithm. This procedure yields a value for the dipole of the different quantities ($\sigma^2$, $\tau^2$, $\chi^2$) for each statistic ($V_0$, $V_1$, $V_2$, and patch variance) for each analysed contrast map\footnote{With the only two exceptions being $\tau^2$ on $V_0$, given that this statistic is independent from $\tau^2$, and the goodness-of-fit $\chi^2$, which cannot be computed directly on the patch. }.

We present the results for the amplitude of the dipoles for each of the three quantities first. Then, we study the possible correlations between them and their clustering. Finally, we study the behaviour of higher multipoles to assess whether they are well characterized by dipoles or they show signs of higher-multipole structure.

\subsection{Dipole amplitude in $\sigma^2$}

\begin{figure}
\includegraphics[width =\linewidth]{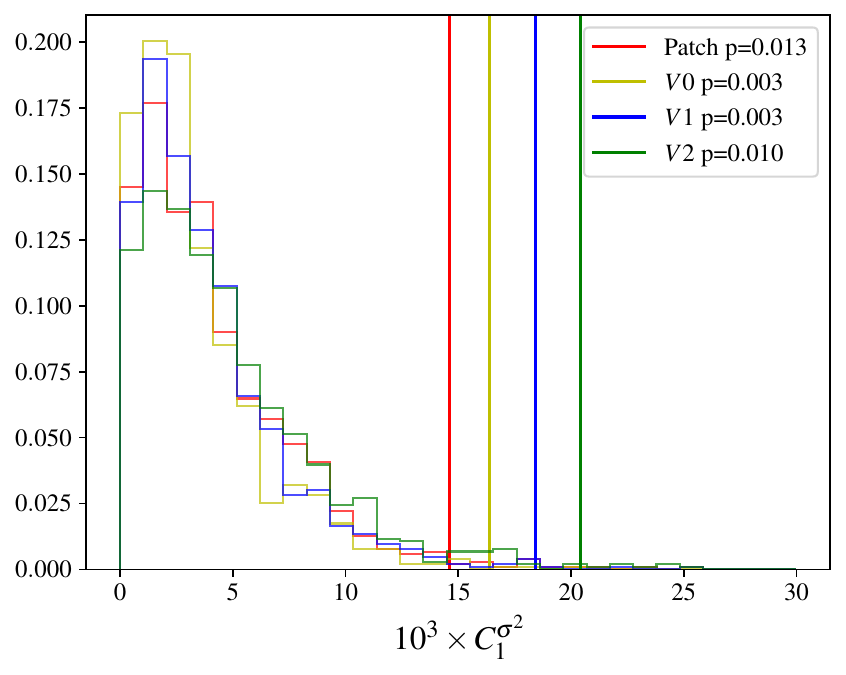}
\caption{Distribution of the amplitude of the dipole in the locally-estimated variance maps, as computed from the 999 realistic CMB simulations, obtained with four different estimators ($V_0$, $V_1$, $V_2$, and patch variance). The vertical lines represent the same quantity in \textit{Planck} CMB data. The $p$-value indicates the fraction of simulations with a dipole amplitude larger than the \textit{Planck} one.}
\label{fig:dipole_significance_sigma}
\end{figure}

We compare the measured amplitude of the dipole on Planck data to the one obtained on simulations, for each MF and the patch variance in \Cref{fig:dipole_significance_sigma}. We can verify that we recover significances similar to those reported in the literature for the dipole when computing the sample variance in patches (red line, $p=1.3\%$). We can further see that all MFs yield a large dipole amplitude, with $p$-values between $0.3\%$ and $1.0\%$. The actual value of the dipole is consistent across different estimators.

Under the null hypothesis of Gaussianity and isotropy, the four quantities are different estimators of the same underlying values (albeit with different noise): the amplitude of the dipole in the variance of the contrast map. Therefore, their values are expected to be highly correlated; in other words, these significances are not independent. In \Cref{ap:corrs} we will study these correlations in detail, but the main point is that these four amplitudes are correlated (with a correlation factor $\sim0.8$), and all the dipoles in Planck data are consistently high, close to the value expected given this correlation. Unlike the dipoles in the following sections, the dipoles in $\sigma^2$ from the MFs can therefore be understood as a consistency check of the patch variance dipole, not as an new independent measurement.

It is worth noting that, when using the standard disc covering requirement of $10\%$, the usual patch variance significance increases, with the $p$-value being reduced to $0.5\%$, in line with other results in the literature, such as \citet{Akrami:2014eta}. We used a more conservative choice of parameters and did not search for choices that maximize the significance. This shows that more significant results can easily be found, and it is not clear whether considering the look-elsewhere effect would decrease or increase the total significance of the anomaly.

\subsection{Dipole amplitude in $\tau^2$}

We repeat the analysis for $\tau^2$, the variance of the spatial derivatives of the field. This parameter can be understood as the correlation length of the field (more technically, its inverse). For Gaussian isotropic fields, this parameter acts as a normalisation factor in the second and third MFs. In this case, we use $V_1$, $V_2$ and patch variance (of the spatial derivatives) as statistics; we cannot use $V_0$ as its theoretical prediction does not depend on $\tau^2$.

\begin{figure}
\includegraphics[width =\linewidth]{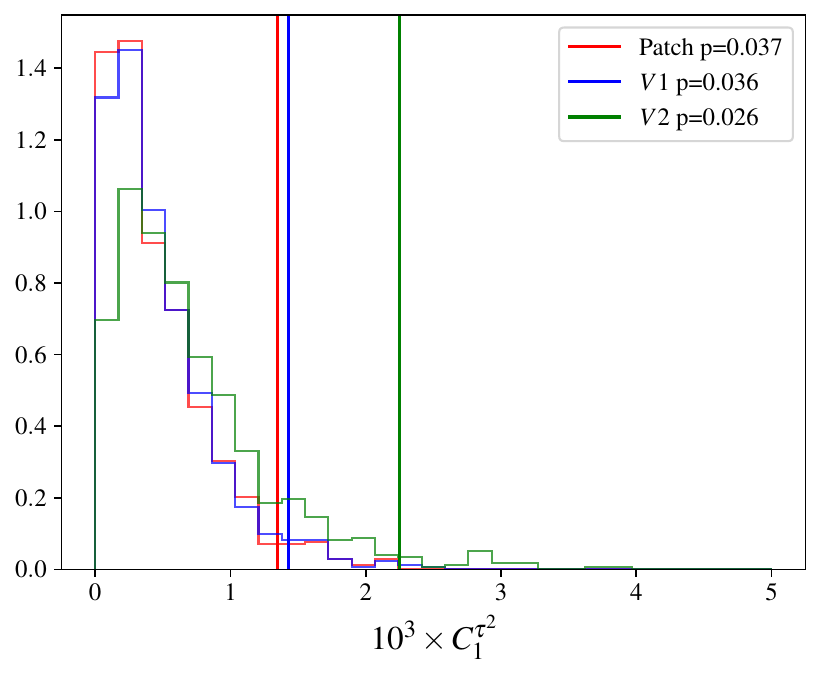}
\caption{Distribution of the amplitude of the dipole in the locally-estimated $\tau^2$ maps, as computed from the 999 realistic CMB simulations, obtained with three different estimators ($V_1$, $V_2$, and patch $\tau^2$). The vertical lines represent the same quantity in \textit{Planck} CMB data. The $p$-value indicates the fraction of simulations with a dipole amplitude larger than the \textit{Planck} one.
}
\label{fig:dipole_significance_tau}
\end{figure}

We show the distribution of the dipole amplitudes for simulations and \textit{Planck} CMB data in  \Cref{fig:dipole_significance_tau}. We see that the $\tau^2$ Planck maps yield moderately high value of the dipole with the three statistics, with $p$-value between $2.6\%$ and $3.7\%$. These $p$-values, albeit not anomalous, seem to be again consistently low, and the dipole amplitude consistently high. These three statistics are again expected to be correlated among themselves (see \Cref{ap:corrs}), as they trace the same underlying property of the field. We will see in \Cref{ssec:indep} that they are, however, independent from the large $\sigma^2$ dipole.

This is compatible with a related result obtained in \cite{Park:2003qd}, where they analysed the amplitude of the genus of excursion sets (proportional to $\tau^2$) in WMAP data and found that the difference between the northern and southern hemispheres was anomalously high ($p$=1.4\%).

\subsection{Deviations from the Gaussian isotropic field predictions}

An important advantage of MFs is that they are very well understood in the mathematical literature, and their expected values are known (with low variance) for Gaussian isotropic fields. This allows us to infer not only the values of $\sigma^2$ and $\tau^2$ of the field, but also how well the MFs actually fit the Gaussian isotropic prediction. We have studied this through the goodness-of-fit, quantified by $\chi^2$ as described in \Cref{sec:data}. Previous studies have shown that the MFs are well fitted by the Gaussian isotropic prediction when analysed as a whole map \citep{Planck:2015igc,Planck:2019evm}. However, to the best of our knowledge, there have been no studies about the uniformity of this fit.

\begin{figure}
\includegraphics[width = \linewidth]{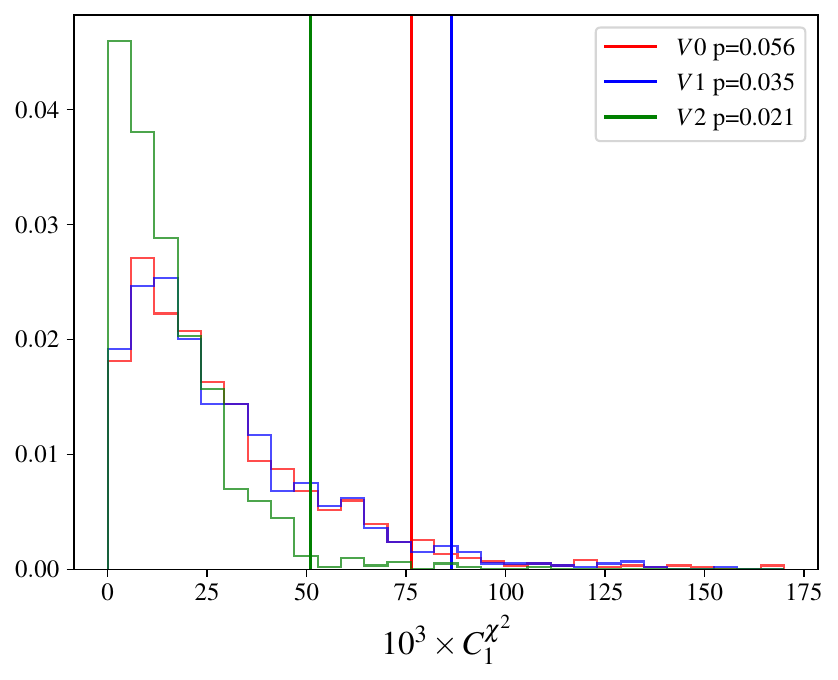}
\caption{Distribution of the amplitude of the dipole in the locally-estimated goodness-of-fit maps, as computed from the 999 realistic CMB simulations, obtained with three different estimators ($V_0$, $V_1$ and $V_2$). The vertical lines represent the same quantity in \textit{Planck} CMB data. The $p$-value indicates the fraction of simulations with a dipole amplitude larger than the \textit{Planck} one.}
\label{fig:GOF_dipole}
\end{figure}

We study the spatial distribution of the goodness-of-fit in the same way as $\sigma^2$ and $\tau^2$. Unlike the previous cases, the $\chi^2$ obtained with the three MFs are not tracing the same underlying quantity and are not necessarily expected to be similar; different kinds of deviations from Gaussianity would be reflected in different ways in the three MFs. A perfect Gaussian isotropic field would follow the predictions for the three MFs. We show the amplitude of the $\chi^2$ dipoles in \Cref{fig:GOF_dipole}, comparing the distribution on the simulations, with the Planck values (vertical lines). We again obtain moderately high values for the dipole amplitudes, with $p$-values of $2.1\%$, $3.5\%$, and $5.6\%$.

\subsection{Independence and alignment of the dipoles}
\label{ssec:indep}

\begin{figure}
    \centering
    \includegraphics[width=\linewidth]{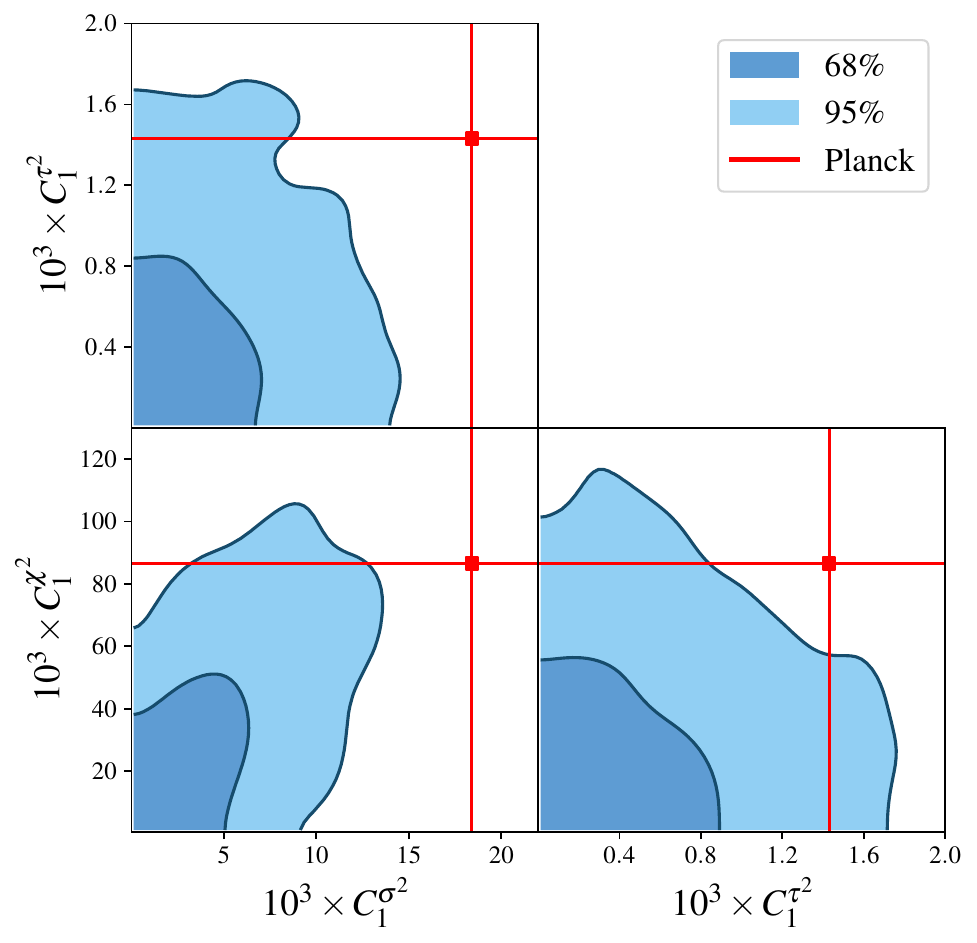}
    \caption{Joint distributions for different pairs of dipole amplitudes extracted from the simulations with the $V_1$ MF statistic. The red lines correspond the dipole amplitudes measured in the \textit{Planck} contrast maps.
    }
    \label{fig:sigmatau_dipole_correlation}
\end{figure}

In the previous sections, we have shown the amplitude of the computed dipoles in $\sigma^2$, $\tau^2$, and $\chi^2$. All of them seem to be unusually high, albeit at different degrees of significance. Two natural questions that arise are whether the ampitudes of these dipoles are expected to be correlated in isotropic $\Lambda$CDM realizations, and whether they point in similar directions.

In \Cref{fig:sigmatau_dipole_correlation}, we show the correlations between the dipole amplitudes for simulations, and compare them to the Planck data, of the $\sigma^2$, $\tau^2$, and $\chi^2$ fields computed with $V_1$. It can be seen that the dipole amplitudes of $\sigma^2$ and $\tau^2$ are uncorrelated (correlation coefficient $r=0.154$), similarly to $\tau^2$ and $\chi^2$ ($r=0.095$). The dipole amplitudes of $\sigma^2$ and $\chi^2$ are, however, moderately correlated ($r=0.733$). Nevertheless, it is interesting to note that the \textit{Planck} dipole amplitudes deviate from the trend present in simulations. This suggests that the mechanism causing the $\sigma^2$ anomaly is likely to be affecting the $\tau^2$ and $\chi^2$ dipoles as well, as these are mostly uncorrelated under the isotropic $\Lambda$CDM model. We obtain similar results for the three dipoles of $V_2$, as well as for the two dipoles of $V_0$ and patch quantities. See \Cref{ap:corrs} for further details.

Joint significance can be estimated in different ways. Notably, no simulation yields a higher value in both $\sigma^2$ and $\tau^2$ at the same time in any of the tested statistics. We also evaluate the combined dipole magnitude, defined as the radial distance from the origin (in units of the \textit{Planck} dipole values). By this metric, only $0.7\%$ of the simulations for $V_1$ and $0.9\%$ for $V_2$ exceed the \textit{Planck} result.

\begin{figure}
\includegraphics[width = \linewidth]{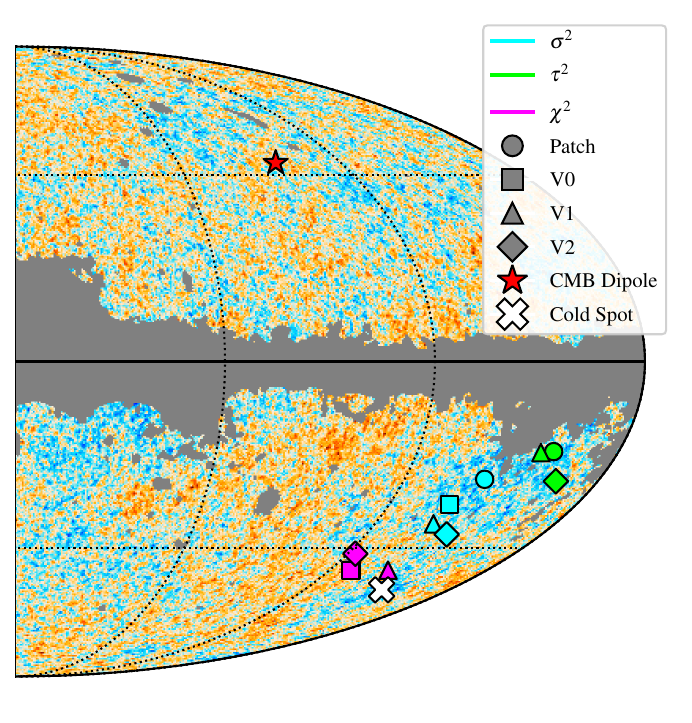}
\caption{Dipole vectors obtained for \textit{Planck} CMB data with different methods. The directions of the CMB dipole and the Cold Spot are also shown.
\label{fig:directions}}
\end{figure}

\begin{figure}
\includegraphics[width = \linewidth]{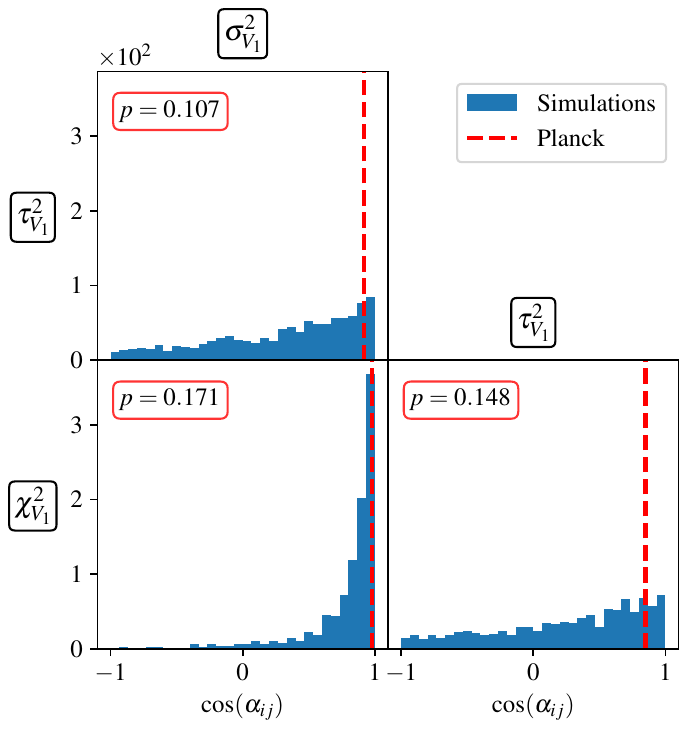}
\caption{Dipole clustering for $V_1$ contrast maps, quantified via the dot product of the dipole directions. The red dashed lines show the dipole distances in the Planck data.
\label{fig:clusteringV1}}
\end{figure}

The other important aspect to investigate is whether these dipoles point towards the same direction and, if so, whether the alignment is expected or unusual. \Cref{fig:directions} shows the direction of all the \textit{Planck} dipoles obtained in this work, together with the CMB dipole. The exact directions can be found in \Cref{tab:dipole_dir}. The first thing to notice is that all of them are aligned, pointing towards the same small region of the sky. The dipoles seem to be grouped in three clusters, one for each computed quantity $\sigma^2$, $\tau^2$, and $\chi^2$; this is expected as they follow the same underlying phenomenon. The three clusters are closer together than expected, although not abnormally so: taking the $V_1$ dipoles as an example, we can see the (cosine of the) angular distance between dipoles in \Cref{fig:clusteringV1}, yielding $p$-values of $10.7\%$, $14.8\%$, and $17.1\%$. Similar values are obtained for all MFs and patch-computed quantities, see \Cref{ap:multipoles} for further details.

We note that these dipoles, especially the ones for $\chi^2$, point very close to the Cold Spot \citep{Vielva:2003et,Cruz:2004ce}. This is an unusually cold region of the CMB, located at $(\ell,b)\approx(208,-56)$, and believed to be caused by the Eridanus Supervoid \citep{Szapudi:2014zha}. The possible relation between the Cold Spot and the dipoles presented here is outside the scope of this work, but it has been claimed before that the $\sigma^2$ dipole could be closely related to the Cold Spot \citep{Inoue:2006rd, Bernui:2009pv}. Some other groups, however, concluded that the Cold Spot was not responsible for the asymmetry \citep{Paci_2010}. We believe that it will be worth studying the effect of the Cold Spot on these new $\tau^2$ and $\chi^2$ dipoles and will undertake such endeavour in an upcoming paper. 

\begin{table}
    \centering
    \begin{tabular}{cc}
        Dipole & Direction $(\ell,\,b)$ \\\hline\hline
$\sigma^2_{v_0}$    &(220.6, -33.9)\\
$\sigma^2_{v_1}$    &(220.5, -38.7)\\
$\sigma^2_{v_2}$    &(212.5, -41.5)\\
$\sigma^2_{patch}$    &(215.3, -27.7)\\ \hline
$\tau^2_{v_1}$    &(203.0, -21.3)\\
$\tau^2_{v_2}$    &(192.9, -28.2)\\
$\tau^2_{patch}$    &(199.4, -21.0)\\\hline
$\chi^2_{v_0}$    &(231.7, -51.1)\\
$\chi^2_{v_1}$    &(217.7, -51.0)\\
$\chi^2_{v_2}$    &(237.3, -46.5)\\
    \end{tabular}
    \caption{Direction of dipole fitted to different maps.}
    \label{tab:dipole_dir}
\end{table}

\subsection{Higher multipole structure}\label{ssec:higher_mult}

In the previous sections, we focused on the dipoles of the different analysed fields ($\sigma^2$, $\tau^2$, and $\chi^2$; with the different MFs and directly on patches). We observe that the dipoles in Planck data are atypical in realistic simulations, at different degrees of significance. A natural question is whether this anomaly is limited to the dipole, or whether it present some structure at higher multipoles. 

\begin{figure}
    \centering
    \includegraphics[width=\linewidth]{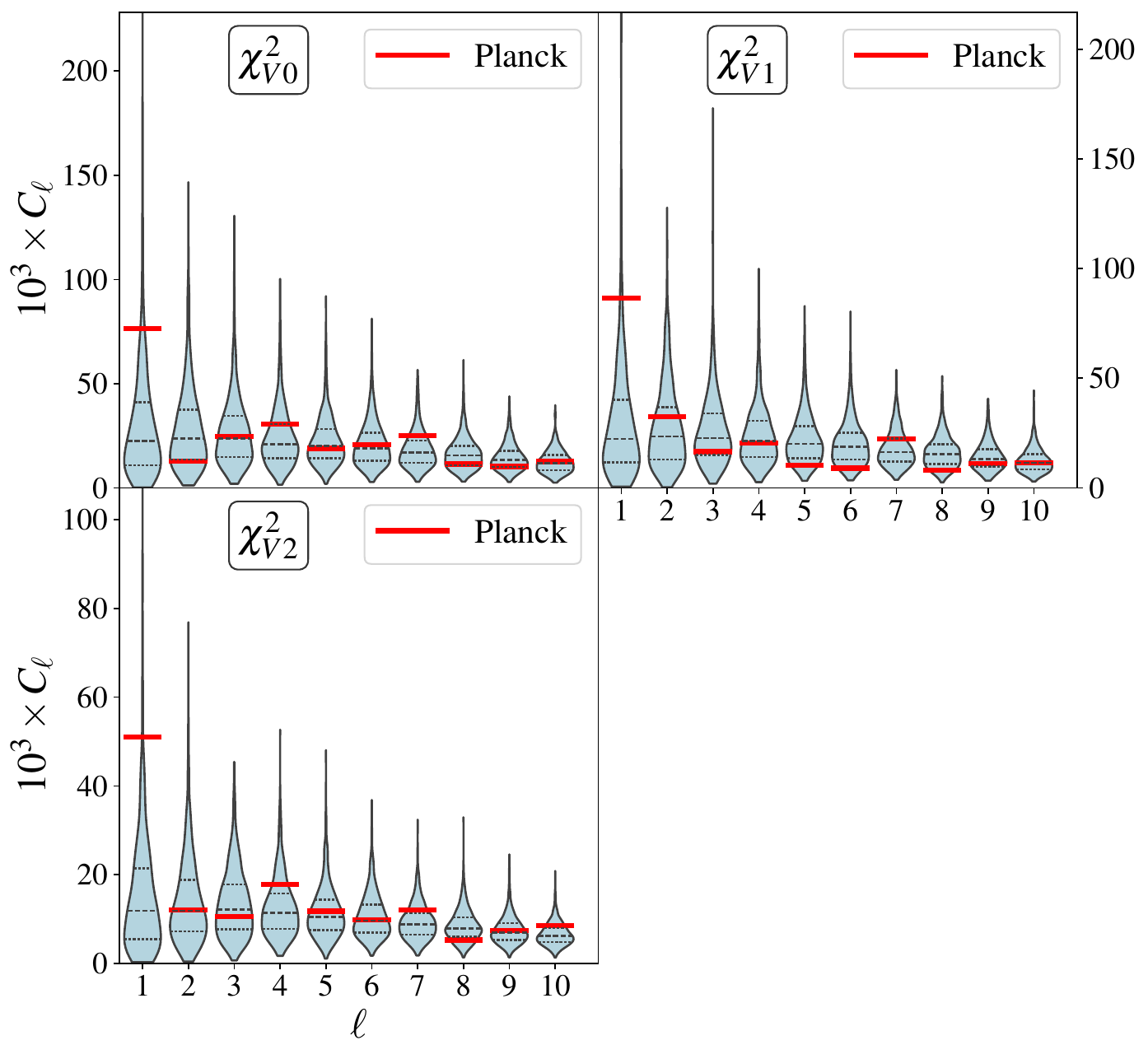}
    \caption{Violin plot representation of the $\chi^2$ contrast maps. The red horizontal tags display the results for \textit{Planck} data. }
    \label{fig:Chi2Multipoles}
\end{figure}

In \Cref{fig:Chi2Multipoles} we show the distribution of the angular power spectrum obtained for $\chi^2$ maps in simulations and Planck data for multipoles up to $\ell=10$. We can see that the dipole is the multipole with the most significant deviation from the simulations. Although the distribution of the Planck values may seem to be slightly more atypical than expected, we note that these multipoles are not fully uncorrelated because of the Planck mask. Therefore, one cannot simply add the significances of different multipoles without a detailed study of their correlations, which is outside of the scope of this work. Similar conclusions are obtained from the $\sigma^2$ and $\tau^2$ angular power spectra, with only the dipole being clearly unusual; the full figure with all these quantities can be found in \Cref{ap:multipoles}.

We can conclude that all the asymmetries studied in this work are well characterized by a dipole. In fact, all of them seem to be compatible with a dipolar effect in the same direction, as we showed in the last section. A more complicated structure at higher multipoles cannot be ruled out, for example if the effect is caused by a local feature, which could be diluted into several multipoles, with a strong dipole being the first-order approximation. However, we find no indication to prefer this explanation versus a pure dipole. This is consistent with our remark at the end of \Cref{sec:MFs}, claiming that the Gaussian isotropic predictions of the MFs should hold on local patches. 

\section{Conclusions}\label{sec:conclusions}

In this work, we have revisited the hemispherical asymmetry of the CMB using new statistics beyond the variance, in particular using MFs. While standard analyses of this anomaly focus primarily on the variance of the temperature field, MFs allow us to probe the morphological and topological properties of the CMB. We analyze the local behaviour of the three MFs ($V_0, V_1, V_2$) on \textit{Planck} SMICA maps and compare them to FFP10 simulations. By contrasting with the theoretical expectation of these statistics, we have characterized the asymmetry in terms of the field variance ($\sigma^2$), the gradient variance ($\tau^2$), and the goodness-of-fit to the Gaussian isotropic prediction ($\chi^2$). To our knowledge, this is the first study on the uniformity of the MFs over the CMB. Our main findings are the following:

\begin{itemize}
    \item We confirm the presence of the hemispherical power asymmetry with high significance. By extracting the local variance $\sigma^2$ from the MFs, we find a dipole amplitude with a $p$-value between $0.3\%$ and $1.0\%$ relative to realistic simulations. This result is consistent with the direct patch-variance estimation, serving as a validation tool for both our pipeline and the presence of the power asymmetry in \textit{Planck} data.
    
    \item We report a moderate dipole in the gradient variance, $\tau^2$, with $p$-values ranging between $2.6\%$ and $3.7\%$. While the statistical significance is lower than that of the $\sigma^2$ dipole, this signal provides non-redundant information: we find that the dipole in $\tau^2$ is statistically independent of the dipole in $\sigma^2$ within the standard $\Lambda$CDM simulations. This implies that the observed asymmetry in the gradients is not merely a byproduct of the power asymmetry; rather, it constitutes an additional constraint that any theoretical model attempting to explain the CMB anomalies must satisfy.
     
    \item We analyzed the spatial distribution of the goodness-of-fit ($\chi^2$) to the Gaussian isotropic predictions. We found a dipole in the quality of the fit with $p$-values between $2.1\%$ and $5.6\%$. Although the significance is low, this hints at a spatial variation in how well the CMB morphology is described by a Gaussian field.

    \item All these dipoles seem to be highly clustered, pointing in a similar direction of the sky (see \Cref{fig:directions}).

    \item These coherent global variations over the sky seem to be well described by a dipole. Although smaller structures cannot be ruled out, the significance of the anomaly of the dipole is much larger than any other single multipole. 
\end{itemize}

While we have restricted our analysis to the SMICA pipeline and temperature data due to computational time limitations, future work could extend this methodology to the other map-making algorithms and polarization maps. The interplay between $E$-mode morphology and temperature anisotropy could further distinguish between a cosmological origin, a systematic effect, or an unexplained foreground. 

These results have important implications for the physical interpretation of the hemispherical asymmetry. Phenomenological models that propose a simple modulation of the primordial power spectrum must be tuned to simultaneously reproduce the asymmetry in the amplitudes of both $\sigma^2$ and $\tau^2$. The hint of anisotropy in the goodness-of-fit to the Gaussian prediction also suggests that the anomaly might be connected to non-Gaussian features. This suggests that the anomaly could be produced by a residual or unidentified galactic foreground, as these typically induce large-scale variations and are naturally non-Gaussian. However, more research is needed before a conclusion can be reached, and an anisotropy of cosmological origin cannot be ruled out. The results in this paper establish new properties of this anomaly that should be satisfied by models that aim to explain it; this makes model building more demanding, but a well-motivated model that explains all aspects of this anomaly would become a strong contestant to explain this effect.

\begin{acknowledgements}
The authors thank Raúl E. Angulo, Yashar Akrami, Arthur Kosowski, Glenn Starkman, and Craig Copi for useful discussions and comments. 
Some of the results in this paper have been derived using the healpy and HEALPix package. 
This work is partially supported by the Spanish Research Agency (Agencia Estatal de Investigaci\'on) through the Grant IFT Centro de Excelencia Severo Ochoa No. CEX2020-001007-S, funded by MCIN/AEI/10.13039/501100011033. 
J.M.A. acknowledges support by the CSIC (Consejo Superior de Investigaciones Científicas) through the JAE Intro programme No. JAEINT-24-01844.
J.C.D. is supported by the Spanish Research Agency (Agencia Estatal de Investigaci\'on), the Ministerio de Ciencia, Inovaci\'on y Universidades, and the European Social Funds through grant JDC2023-052152-I, as part of the Juan de la Cierva program. 
M.M.B. acknowledges support by the Spanish Ministry of Science, Innovation and Universities under the FPU predoctoral grant FPU22/02306. 
This work is partially funded by the European Commission – NextGenerationEU, through Momentum CSIC Programme: Develop Your Digital Talent. Numerical calculations have been performed on the Hydra cluster at IFT. We acknowledge HPC support by Emilio Ambite, staff hired under the Generation D initiative, promoted by Red.es, an organisation attached to the Spanish Ministry for Digital Transformation and the Civil Service, for the attraction and retention of talent through grants and training contracts, financed by the Recovery, Transformation and Resilience Plan through the European Union’s Next Generation funds.
\end{acknowledgements}
\bibliographystyle{aa}
\bibliography{refs} 

\appendix

\section{Correlations}
\label{ap:corrs}

In this appendix we explore correlations between different estimators of the physical parameters analyzed in this work. In particular, we are interested in how estimators for the same underlying quantity compare, as well as correlations among estimators for different quantities.

\begin{figure*}
\centering
   \includegraphics[width=0.85\linewidth]{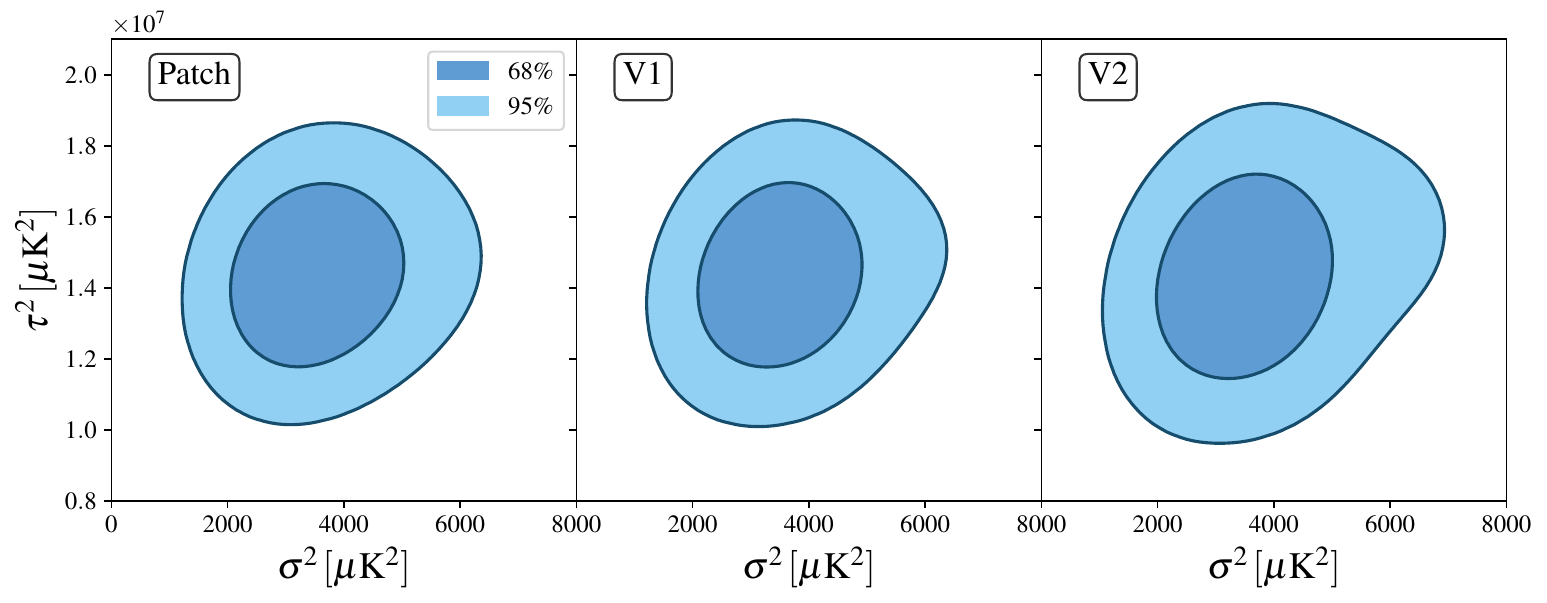}
     \caption{Scatter distributions comparing the local estimators of $\sigma^2$ and $\tau^2$ evaluated across all the valid discs in all the simulations. There is no $V_0$ panel, given that the theoretical prediction for that MF is independent of $\tau^2$.}
     \label{fig:sigma_tau_correlation}
\end{figure*}

Firstly, we explore whether there exists any correlation between the inferred $\sigma^2$ and $\tau^2$ quantities at the disc level. \Cref{fig:sigma_tau_correlation} shows the scatter of these two parameters evaluated across all valid local patches in the simulation suite. As anticipated in \Cref{sec:data}, the local evaluations of $\sigma^2$ and $\tau^2$ are largely uncorrelated. 
Their mutual independence at the patch level confirms that our morphological descriptors are capturing distinct statistical properties of the temperature field.

\begin{figure*}
\centering
   \includegraphics[width=0.85\linewidth]{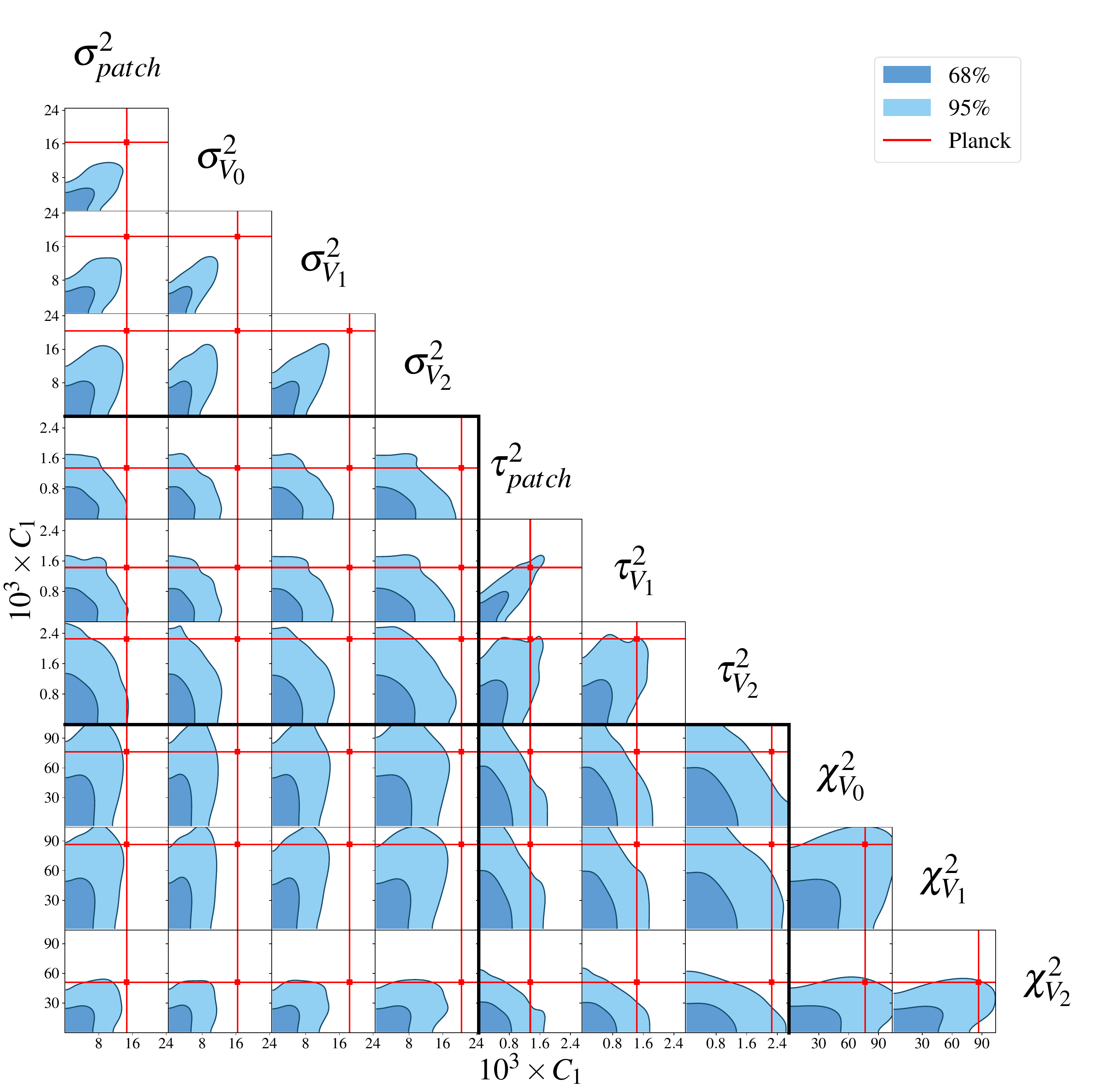}
     \caption{Joint distributions for different pairs of dipole amplitudes extracted from the simulations with all the statistics used in this work. The red lines correspond the dipole amplitudes measured in the \textit{Planck} maps.}
     \label{fig:big_dipamp}
\end{figure*}

Next, we check how correlated the dipole amplitudes of the contrast maps associated to different estimators and quantities are. The corner plot in \Cref{fig:big_dipamp} shows the joint distributions of these amplitudes derived from the FFP10 simulations. As expected, estimators that trace the same underlying physical property exhibit strong positive correlations. In contrast, the dipole amplitudes of distinctly different physical quantities, particularly $\sigma^2$ and $\tau^2$, display minimal correlation under the standard isotropic $\Lambda$CDM model.

\begin{figure*}
\centering
   \includegraphics[width=1\linewidth]{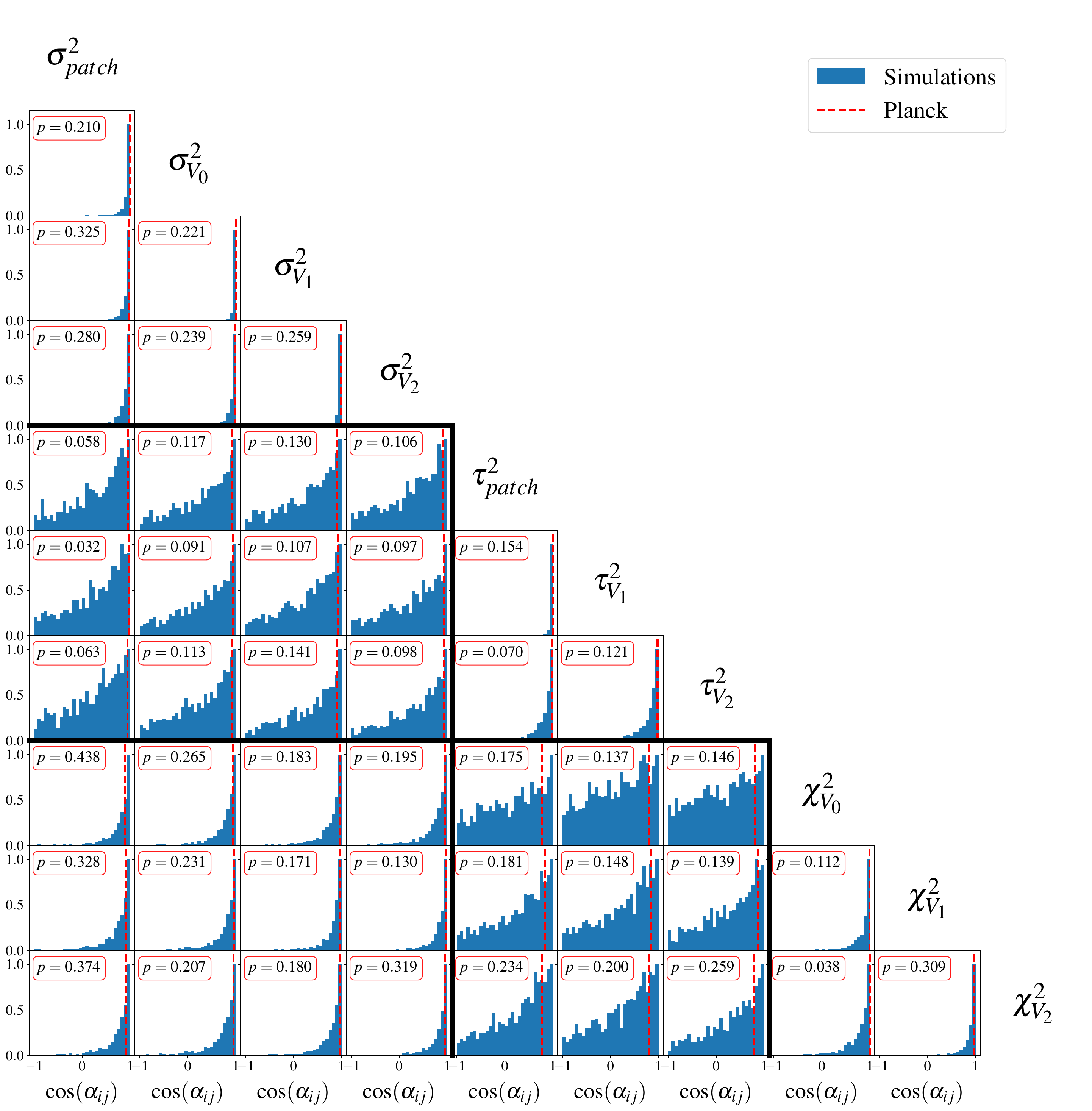}
     \caption{Dipole clustering for all combinations of contrast maps, quantified via the dot product of the dipole directions. The red dashed lines show the dipole distances in the Planck data. The fraction of simulations with stronger alignment than the one observed in \textit{Planck} is reported within the red boxes.}
     \label{fig:big_cosine_plot}
\end{figure*}

\begin{figure*}
\centering
\includegraphics[width =0.65\linewidth]{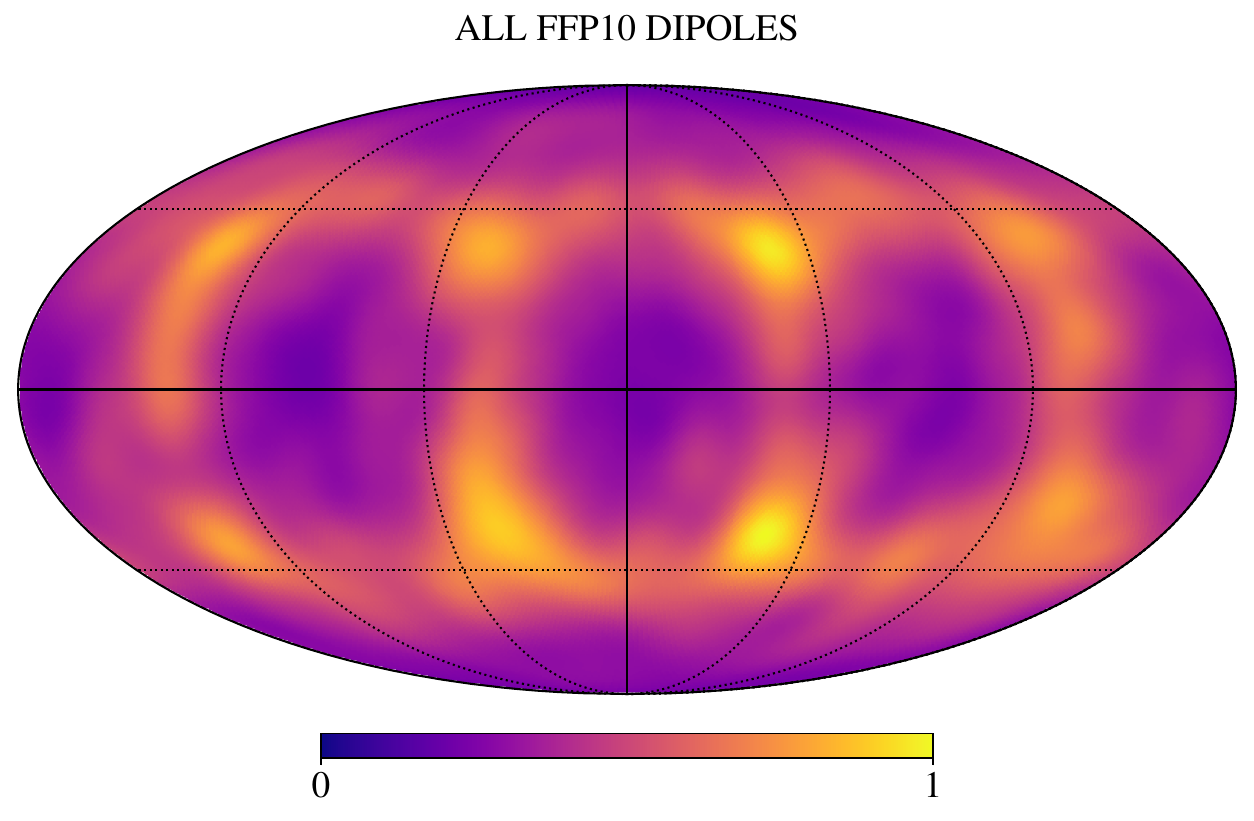}
\caption{Smoothed density distribution of all dipole directions obtained from all different estimators and quantities. The fact that the distribution is not uniform is a consequence of the presence of the mask.}\label{fig:heatmap}
\end{figure*}

Finally, we expand on the dipole alignments discussed in \Cref{ssec:indep}. \Cref{fig:big_cosine_plot} provides the complete set of mutual clustering distributions, quantified via the dot product of the dipole directions, showing how the directional alignments of the various dipoles compare between the Planck data and the simulation suite. Complementarily, \Cref{fig:heatmap} depicts the heatmap of all dipole directions obtained from simulations. The fact that there seem to be preferred directions following a quadrupolar pattern is merely a consequence of the \textit{Planck} mask, and it shows that even isotropic simulations will show a certain degree of clustering.

\newpage
\section{Higher Multipoles}
\label{ap:multipoles}

In \Cref{ssec:higher_mult}, we established that the spatial distribution of the goodness-of-fit $(\chi^2)$ is well characterized by a dipole, lacking highly anomalous structure at higher multipoles. In this appendix, we extend this analysis to the field variance $(\sigma^2)$ and the gradient variance $(\tau^2)$.

\begin{figure*}
\centering
   \includegraphics[width=0.8\linewidth]{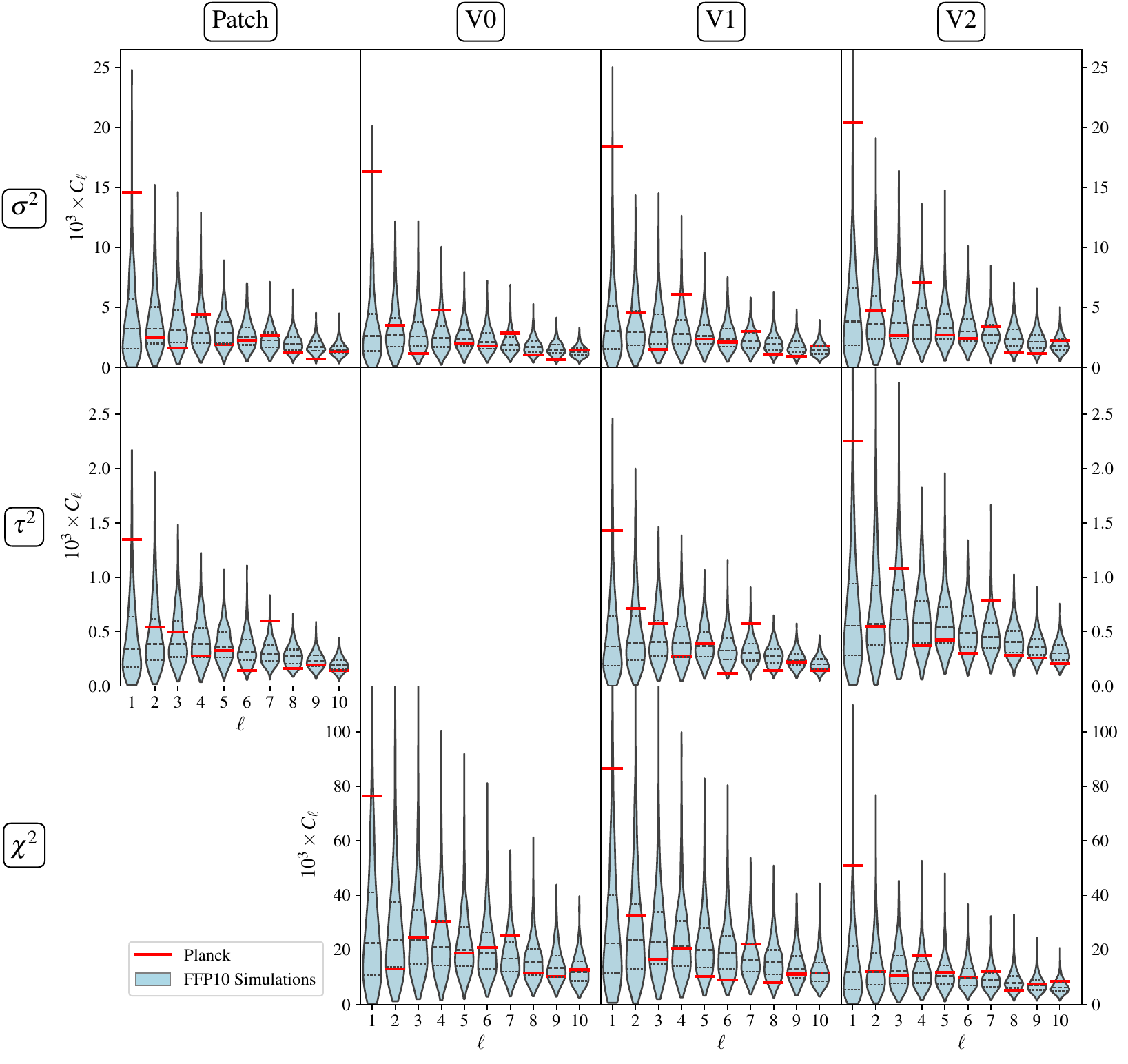}
     \caption{Violin plot representations of the angular power spectra $C_\ell$ up to $\ell=10$ for the three quantities of interest $(\sigma^2,\tau^2,\chi^2)$ using the four different statistics employed in the paper: patch, $V_0,V_1,V_2$. The blue distributions are computed from the simulations, while the red lines correspond to the angular power spectrum measured in the \textit{Planck} maps.}
     \label{fig:big_higher_mult}
\end{figure*}

We calculate the angular power spectra of the contrast maps for all estimators up to $\ell=10$. The distributions of these multipoles for the FFP10 simulations, alongside the measured Planck values, are shown in \Cref{fig:big_higher_mult}. As anticipated in the main body of the text, we can see that the dipole is the only multipole that is anomalously high on its own.

\end{document}